# Wireless Secrecy in Large-Scale Networks

Pedro C. Pinto, *Member, IEEE*, João Barros, *Member, IEEE*,
and Moe Z. Win, *Fellow, IEEE*

*Abstract*—The ability to exchange secret information is critical to many commercial, governmental, and military networks. The *intrinsically secure communications graph* ($i\mathcal{S}$-graph) is a random graph which describes the connections that can be securely established over a large-scale network, by exploiting the physical properties of the wireless medium. This paper provides an overview of the main properties of this new class of random graphs. We first analyze the *local properties* of the $i\mathcal{S}$-graph, namely the degree distributions and their dependence on fading, target secrecy rate, and eavesdropper collusion. To mitigate the effect of the eavesdroppers, we propose two techniques that improve secure connectivity. Then, we analyze the *global properties* of the $i\mathcal{S}$-graph, namely percolation on the infinite plane, and full connectivity on a finite region. These results help clarify how the presence of eavesdroppers can compromise secure communication in a large-scale network.

## I. INTRODUCTION

Contemporary security systems for wireless networks are based on cryptographic primitives that generally ignore two key factors: (a) the physical properties of the wireless medium, and (b) the spatial configuration of both the legitimate and malicious nodes. These two factors are important since they affect the communication channels between the nodes, which in turn determine the fundamental secrecy limits of a wireless network. In fact, the inherent randomness of the wireless medium and the spatial location of the nodes can be leveraged to provide *intrinsic security* of the communications infrastructure at the physical-layer level.[1]

The basis for information-theoretic security, which builds on the notion of perfect secrecy [1], was laid in [2] and later in [3], [4]. More recently, there has been a renewed interest in information-theoretic security over wireless channels, from the perspective of space-time communications [5], multiple-input multiple-output communications [6]–[10], eavesdropper collusion [11], [12], cooperative relay networks [13], fading channels [14]–[18], strong secrecy [19], [20], secret key agreement [21]–[25], code design [26]–[28], among other topics.

P. C. Pinto is with the Swiss Federal Institute of Technology (EPFL), Lausanne, Switzerland (e-mail: pedro.pinto@epfl.ch). J. Barros is with Faculdade de Engenharia da Universidade do Porto (FEUP), Porto, Portugal (e-mail: jbarros@fe.up.pt). M. Z. Win is with the Massachusetts Institute of Technology, Cambridge, MA, USA (e-mail: moewin@mit.edu).

This research was supported, in part, by the Portuguese Science and Technology Foundation under grant SFRH-BD-17388-2004; the MIT Institute for Soldier Nanotechnologies; the Office of Naval Research under Presidential Early Career Award for Scientists and Engineers (PECASE) N00014-09-1-0435; and the National Science Foundation under grant ECS-0636519.

[1]In the literature, the term "security" typically encompasses 3 different characteristics: *secrecy* (or privacy), *integrity*, and *authenticity*. This paper does not consider the issues of integrity or authenticity, and the terms "secrecy" and "security" are used interchangeably.

A comprehensive treatment of physical-layer security can be found in [29]. A fundamental limitation of the literature is that it only considers scenarios with a small number of nodes. To account for large-scale networks composed of multiple legitimate and eavesdropper nodes, *secrecy graphs* were introduced in [30] from a geometrical perspective, and in [31] from an information-theoretic perspective. The local connectivity of secrecy graphs was extensively characterized in [32], while the scaling laws of the secrecy capacity were presented in [33], [34]. The feasibility of long-range secure communication was proved in [35], in the context of continuum percolation.

In this paper, we present an overview of secure communication over large-scale networks, in terms of the properties of the underlying random graph. The main contributions are as follows:

- *Framework for intrinsic security in stochastic networks:* We introduce an information-theoretic definition of the intrinsically secure communications graph ($i\mathcal{S}$-graph), based on the notion of strong secrecy.
- *Local connectivity in the $i\mathcal{S}$-graph:* We provide a probabilistic characterization of both in-degree and out-degree of a typical node.
- *Techniques for communication with enhanced secrecy:* We propose sectorized transmission and eavesdropper neutralization as two techniques for enhancing the secrecy of communication.
- *Maximum secrecy rate (MSR) in the $i\mathcal{S}$-graph:* We provide a probabilistic characterization of the MSR between a typical legitimate node and each of its neighbors.
- *The case of colluding eavesdroppers:* We quantify the degradation in secure connectivity arising from eavesdroppers collusion.
- *Percolation in the $i\mathcal{S}$-graph:* We prove the existence of a phase transition in the Poisson $i\mathcal{S}$-graph, showing that long-range communication is still possible when a secrecy constraint is present.
- *Full connectivity in the $i\mathcal{S}$-graph:* We characterize secure full connectivity on a finite region of the Poisson $i\mathcal{S}$-graph.

This paper is organized as follows. Section II describes the system model. Section III characterizes local connectivity in the Poisson $i\mathcal{S}$-graph. Section IV analyzes two techniques for enhancing the secrecy of communication. Section V considers the MSR between a node and its neighbours. Section VI characterizes the case of colluding eavesdroppers. Section VII characterizes continuum percolation in the Poisson $i\mathcal{S}$-graph defined over the infinite plane. Section VIII analyzes full connectivity in the Poisson $i\mathcal{S}$-graph restricted to a finite region. Section IX concludes the paper.

## II. SYSTEM MODEL

### A. Wireless Propagation Characteristics

In a wireless environment, the received power $P_{\text{rx}}(x_i, x_j)$ associated with the link $\overrightarrow{x_i x_j}$ can modeled as

$$P_{\text{rx}}(x_i, x_j) = P_\ell \cdot g(x_i, x_j, Z_{x_i,x_j}), \quad (1)$$

where $P_\ell$ is the (common) transmit power of the legitimate nodes; and $g(x_i, x_j, Z_{x_i,x_j})$ is the power gain of the link $\overrightarrow{x_i x_j}$, where the random variable (RV) $Z_{x_i,x_j}$ represents the random propagation effects (such as multipath fading or shadowing) associated with link $\overrightarrow{x_i x_j}$. We consider that the $Z_{x_i,x_j}, x_i \neq x_j$ are independent identically distributed (IID) RVs with common probability density function (PDF) $f_Z(z)$, and that $Z_{x_i,x_j} = Z_{x_j,x_i}$ due to channel reciprocity. The channel gain $g(x_i, x_j, Z_{x_i,x_j})$ is considered constant (quasi-static) throughout the use of the communications channel, which corresponds to channels with a large coherence time. The gain function is assumed to satisfy the following conditions:

1) $g(x_i, x_j, Z_{x_i,x_j})$ depends on $x_i$ and $x_j$ only through the link length $|x_i - x_j|$; with abuse of notation, we can write $g(r, z) \triangleq g(x_i, x_j, z)|_{|x_i - x_j| \to r}$.[2]
2) $g(r, z)$ is continuous and strictly decreasing in $r$.
3) $\lim_{r \to \infty} g(r, z) = 0$.

The proposed model is general enough to account for common choices of $g$. One example is the unbounded model where $g(r,z) = \frac{z}{r^{2b}}$. The term $\frac{1}{r^{2b}}$ accounts for the far-field path loss with distance, where the amplitude loss exponent $b$ is environment-dependent and can approximately range from 0.8 (e.g., hallways inside buildings) to 4 (e.g., dense urban environments), with $b = 1$ corresponding to free space propagation. Another example is the bounded model where $g(r,z) = \frac{z}{1+r^{2b}}$, which eliminates the singularity at the origin, but often leads to intractable analytical results [36]. Furthermore, by appropriately choosing of the distribution of $Z_{x_i,x_j}$, both models can account for various random propagation effects, including Nakagami-$m$ fading, Rayleigh fading, and log-normal shadowing [37].

### B. $i\mathcal{S}$-Graph

Consider a wireless network where legitimate nodes and potential eavesdroppers are randomly scattered in space, according to some point process. The $i\mathcal{S}$-graph is a convenient representation of the information-theoretically secure links that can be established on such network, and is defined as follows.

*Definition 2.1 ($i\mathcal{S}$-Graph [31]):* Let $\Pi_\ell = \{x_i\} \subset \mathbb{R}^d$ denote the set of legitimate nodes, and $\Pi_e = \{e_i\} \subset \mathbb{R}^d$ denote the set of eavesdroppers. The $i\mathcal{S}$-graph is the directed graph $G = \{\Pi_\ell, \mathcal{E}\}$ with vertex set $\Pi_\ell$ and edge set

$$\mathcal{E} = \{\overrightarrow{x_i x_j} : \mathcal{R}_\mathsf{s}(x_i, x_j) > \varrho\}, \quad (2)$$

where $\varrho$ is a threshold representing the prescribed infimum secrecy rate for each communication link; and $\mathcal{R}_\mathsf{s}(x_i, x_j)$ is

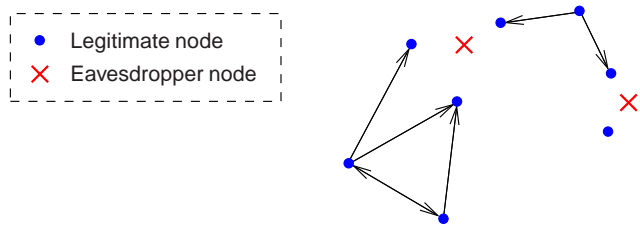

Figure 1. Example of an $i\mathcal{S}$-graph on $\mathbb{R}^2$.

the *maximum secrecy rate* (MSR) of the link $\overrightarrow{x_i x_j}$, given by

$$\mathcal{R}_\mathsf{s}(x_i, x_j) = \left[\log_2\left(1 + \frac{P_{\text{rx}}(x_i, x_j)}{\sigma_\ell^2}\right) - \log_2\left(1 + \frac{P_{\text{rx}}(x_i, e^*)}{\sigma_\mathsf{e}^2}\right)\right]^+ \quad (3)$$

in bits per complex dimension, where $[x]^+ = \max\{x, 0\}$; $\sigma_\ell^2, \sigma_\mathsf{e}^2$ are the noise powers of the legitimate users and eavesdroppers, respectively; and $e^* = \operatorname{argmax}_{e_k \in \Pi_\mathsf{e}} P_{\text{rx}}(x_i, e_k)$.[3]

This definition presupposes that the eavesdroppers are not allowed to *collude* (i.e., they cannot exchange or combine information), and therefore only the eavesdropper with the strongest received signal from $x_i$ determines the MSR between $x_i$ and $x_j$.

The $i\mathcal{S}$-graph admits an outage interpretation, in the sense that legitimate nodes set a target secrecy rate $\varrho$ at which they transmit without knowing the channel state information (CSI) of the legitimate nodes and eavesdroppers. In this context, an edge between two nodes signifies that the corresponding channel is not in secrecy outage.

Consider now the particular scenario where the following conditions hold: (a) the infimum desired secrecy rate is zero, i.e., $\varrho = 0$; (b) the wireless environment introduces only path loss, i.e., $Z_{x_i,x_j} = 1$ in (1); and (c) the noise powers of the legitimate users and eavesdroppers are equal, i.e., $\sigma_\ell^2 = \sigma_\mathsf{e}^2 = \sigma^2$. Note that by setting $\varrho = 0$, we are considering the *existence* of secure links, in the sense that an edge $\overrightarrow{x_i x_j}$ is present if and only if $\mathcal{R}_\mathsf{s}(x_i, x_j) > 0$. Under these special conditions, the edge set in (2) simplifies to

$$\mathcal{E} = \left\{\overrightarrow{x_i x_j} : |x_i - x_j| < |x_i - e^*|, \quad e^* = \operatorname*{argmin}_{e_k \in \Pi_\mathsf{e}} |x_i - e_k|\right\}, \quad (4)$$

which corresponds the geometrical model proposed in [30]. Fig. 1 shows an example of such an $i\mathcal{S}$-graph on $\mathbb{R}^2$.

The spatial location of the legitimate and eavesdropper nodes can be modeled either deterministically or stochastically. In many cases, the node positions are unknown to the network designer a priori, so they may be treated as uniformly random according to a Poisson point process [37]–[40].

*Definition 2.2 (Poisson $i\mathcal{S}$-graph):* The *Poisson $i\mathcal{S}$-graph* is an $i\mathcal{S}$-graph where $\Pi_\ell, \Pi_\mathsf{e} \subset \mathbb{R}^d$ are mutually independent, homogeneous Poisson point processes with densities $\lambda_\ell$ and $\lambda_\mathsf{e}$, respectively.

In the remainder of the paper (unless otherwise indicated), we focus on Poisson $i\mathcal{S}$-graphs in $\mathbb{R}^2$.

---

[2]For notational simplicity, when $Z = 1$, we omit the second argument of the function $g(r, z)$ and simply use $g(r)$.

[3]This definition uses *strong secrecy* as the condition for information-theoretic security. See [19], [32] for more details.

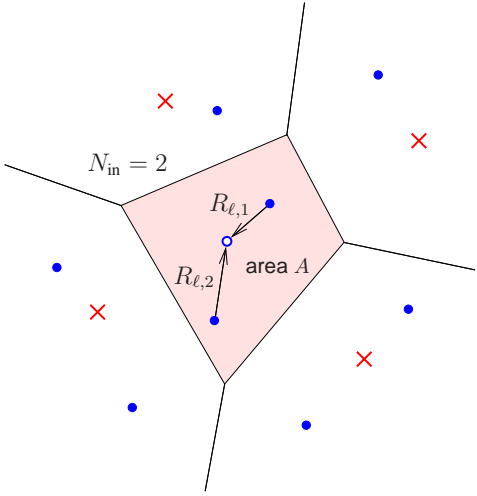

Figure 2. In-degree of a node. In this example, the node at the origin can receive messages with information-theoretic security from $N_{\text{in}} = 2$ nodes. The RV $A$ is the area of a typical Voronoi cell, induced by the eavesdropper Poisson process $\Pi_{\text{e}}$ with density $\lambda_{\text{e}}$.

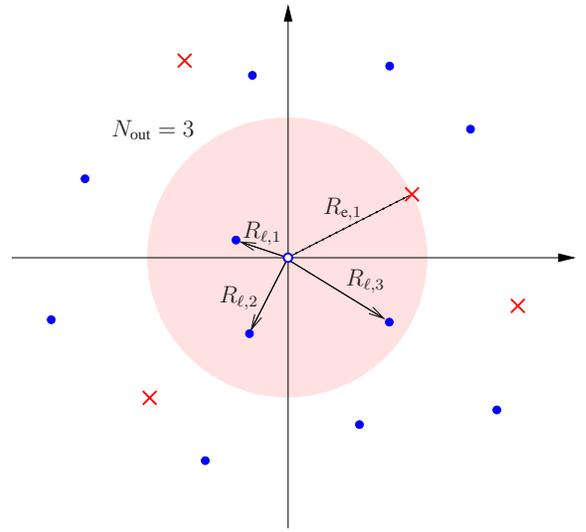

Figure 3. Out-degree of a node. In this example, the node at the origin can transmit messages with information-theoretic security to $N_{\text{out}} = 3$ nodes.

## III. LOCAL CONNECTIVITY IN THE POISSON $i\mathcal{S}$-GRAPH

The $i\mathcal{S}$-graph is a random graph, and therefore the in- and out-degrees of the legitimate nodes are RVs. In this section, we provide a probabilistic characterization of both in-degree $N_{\text{in}}$ and out-degree $N_{\text{out}}$ of a typical node in the Poisson $i\mathcal{S}$-graph.[4] We first consider the simplest case of $\varrho = 0$ (the *existence* of secure links), $Z_{x_i,x_j} = 1$ (path loss only), and $\sigma_{\text{e}}^2 = \sigma_{\ell}^2$ (equal noise powers) in Sections III-A, III-B, and III-C. This scenario leads to an $i\mathcal{S}$-graph with a simple geometric description, thus providing various insights that are useful in understanding more complex cases. Later, in Sections III-D and III-E, we separately analyze how the node degrees are affected by wireless propagation effects other than path loss (e.g., multipath fading), a non-zero secrecy rate threshold $\varrho$, and unequal noise powers $\sigma_{\text{e}}^2, \sigma_{\ell}^2$.

### A. In-Degree Characterization

The following theorem uncovers a surprising connection between a node's in-degree and the area of a typical cell in a Poisson-Voronoi tessellation.

*Theorem 3.1 ([31]):* The in-degree $N_{\text{in}}$ of a typical node in the Poisson $i\mathcal{S}$-graph has the following moment generating function (MGF)

$$M_{N_{\text{in}}}(s) = \mathbb{E}\left\{\exp\left(\frac{\lambda_{\ell}}{\lambda_{\text{e}}}\widetilde{A}(e^s - 1)\right)\right\}, \qquad (5)$$

where $\widetilde{A}$ is the area of a typical Voronoi cell induced by a unit-density Poisson process. Furthermore, all the moments of $N_{\text{in}}$ are given by

$$\mathbb{E}\{N_{\text{in}}^n\} = \sum_{k=1}^{n}\left(\frac{\lambda_{\ell}}{\lambda_{\text{e}}}\right)^k S(n,k)\,\mathbb{E}\{\widetilde{A}^k\}, \quad n \geq 1, \qquad (6)$$

[4] In this paper, we analyze the local properties of a *typical node* in the Poisson $i\mathcal{S}$-graph. This notion is made precise in [39, Sec. 4.4] using Palm theory.

where $S(n,k)$, $1 \leq k \leq n$, are the Stirling numbers of the second kind [41, Ch. 24].

Figure 2 illustrates the in-degree of a legitimate node. Equation (6) expresses the moments of $N_{\text{in}}$ in terms of the moments of $\widetilde{A}$. In general, $\mathbb{E}\{\widetilde{A}^k\}$ cannot be obtained in closed form, except in the case of $k = 1$ where $\mathbb{E}\{\widetilde{A}\} = 1$. For $k = 2$ and $k = 3$, $\mathbb{E}\{\widetilde{A}^k\}$ can be expressed as multiple integrals and then computed numerically [42]. Alternatively, the moments of $\widetilde{A}$ can be determined using Monte Carlo simulation of random Poisson-Voronoi tessellations [43].

The above theorem can be used to obtain other in-connectivity properties, as given in the following corollary.

*Corollary 3.1:* The average in-degree of a typical node in the Poisson $i\mathcal{S}$-graph is

$$\mathbb{E}\{N_{\text{in}}\} = \frac{\lambda_{\ell}}{\lambda_{\text{e}}} \qquad (7)$$

and the probability that a typical node cannot receive from anyone with positive secrecy rate (in-isolation) is

$$p_{\text{in-isol}} = \mathbb{E}\left\{e^{-\frac{\lambda_{\ell}}{\lambda_{\text{e}}}\widetilde{A}}\right\}. \qquad (8)$$

### B. Out-Degree Characterization

*Theorem 3.2 ([30], [31]):* The out-degree $N_{\text{out}}$ of a typical node in the Poisson $i\mathcal{S}$-graph has the following geometric probability mass function (PMF)

$$p_{N_{\text{out}}}(n) = \left(\frac{\lambda_{\ell}}{\lambda_{\ell} + \lambda_{\text{e}}}\right)^n\left(\frac{\lambda_{\text{e}}}{\lambda_{\ell} + \lambda_{\text{e}}}\right), \quad n \geq 0. \qquad (9)$$

Figure 3 illustrates the out-degree of a node. The above theorem can be used to obtain other out-connectivity properties, as given in the following corollary.

*Corollary 3.2:* The average out-degree of a typical node in the Poisson $i\mathcal{S}$-graph is

$$\mathbb{E}\{N_{\text{out}}\} = \frac{\lambda_{\ell}}{\lambda_{\text{e}}}, \qquad (10)$$

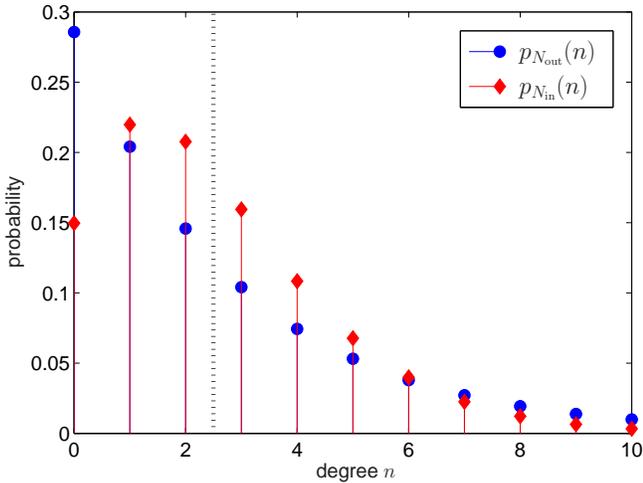

Figure 4. PMF of the in- and out-degree of a node ($\frac{\lambda_e}{\lambda_\ell} = 0.4$). The vertical line marks the average node degrees, $\mathbb{E}\{N_{\text{out}}\} = \mathbb{E}\{N_{\text{in}}\} = \frac{\lambda_\ell}{\lambda_e} = 2.5$, in accordance with Property 3.1.

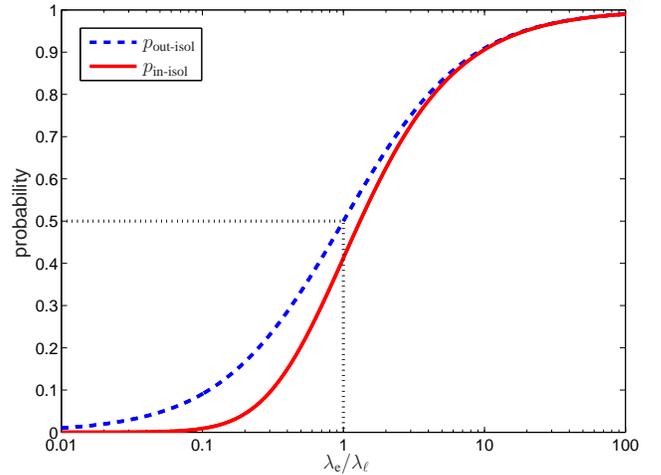

Figure 5. Probabilities of in- and out-isolation of a node, versus the ratio $\frac{\lambda_e}{\lambda_\ell}$. Note that $p_{\text{in-isol}} < p_{\text{out-isol}}$ for any fixed $\frac{\lambda_e}{\lambda_\ell}$, according to Property 3.2.

and the probability that a typical node cannot transmit to anyone with positive secrecy rate (out-isolation) is

$$p_{\text{out-isol}} = \frac{\lambda_e}{\lambda_\ell + \lambda_e}. \quad (11)$$

### C. General Relationships Between In- and Out-Degree

We have so far considered the probabilistic distribution of the in- and out-degrees in a separate fashion. This section establishes a direct comparison between some characteristics of the in- and out-degrees.

*Property 3.1 ([32]):* For the Poisson $i\mathcal{S}$-graph with $\lambda_\ell > 0$ and $\lambda_e > 0$, the average degrees of a typical node satisfy

$$\mathbb{E}\{N_{\text{in}}\} = \mathbb{E}\{N_{\text{out}}\} = \frac{\lambda_\ell}{\lambda_e}. \quad (12)$$

Furthermore, we can establish the following relationship between the probabilities of in- and out-isolation.

*Property 3.2 ([32]):* For the Poisson $i\mathcal{S}$-graph with $\lambda_\ell > 0$ and $\lambda_e > 0$, the probabilities of in- and out-isolation of a typical node satisfy

$$p_{\text{in-isol}} < p_{\text{out-isol}}. \quad (13)$$

An intuitive explanation for this property is provided in [32]. Figure 4 compares the PMFs of the in- and out-degree of a node, while Figure 5 compares the probabilities of in- and out-isolation for various ratios $\frac{\lambda_e}{\lambda_\ell}$.

### D. Effect of the Wireless Propagation Characteristics

We have so far analyzed the local connectivity of the $i\mathcal{S}$-graph in the presence of path loss only. However, the wireless medium typically introduces random propagation effects such as multipath fading and shadowing, which are modeled by the RV $Z_{x_i,x_j}$ in (1). Considering $\varrho = 0$, $\sigma_\ell^2 = \sigma_e^2 = \sigma^2$, and arbitrary propagation effects $Z_{x_i,x_j}$ with PDF $f_Z(z)$, we can combine (2) with the general propagation model of (1) to obtain the edge set

$$\mathcal{E} = \{\overrightarrow{x_i x_j} : g(|x_i - x_j|, Z_{x_i,x_j}) > g(|x_i - e^*|, Z_{x_i,e^*})\}, \quad (14)$$

where

$$e^* = \underset{e_k \in \Pi_e}{\arg\max}\, g(|x_i - e_k|, Z_{x_i,e_k}).$$

Unlike the case of path-loss only, where the out-connections of a node are determined only by the *closest* eavesdropper, here they are determined by the eavesdropper with the *least attenuated* channel. The following theorem characterizes the distribution of the out-degree.

*Theorem 3.3 ([32]):* For the Poisson $i\mathcal{S}$-graph with propagation effects $Z_{x_i,x_j}$ whose PDF is given by a continuous function $f_Z(z)$, the PMF of the out-degree $N_{\text{out}}$ of a typical node is given in (9), and is *invariant* with respect to $f_Z(z)$.

Intuitively, the propagation environment affects both the legitimate nodes and eavesdroppers in the same way, such that the PMF of $N_{\text{out}}$ is invariant with respect to the PDF $f_Z(z)$. However, the PMF of $N_{\text{in}}$ *does* depend on $f_Z(z)$ in a non-trivial way, although its mean remains the same, as specified in the following corollary.

*Corollary 3.3:* For the Poisson $i\mathcal{S}$-graph with propagation effects $Z_{x_i,x_j}$ distributed according to $f_Z(z)$, the average node degrees are

$$\mathbb{E}\{N_{\text{in}}\} = \mathbb{E}\{N_{\text{out}}\} = \frac{\lambda_\ell}{\lambda_e}, \quad (15)$$

for any distribution $f_Z(z)$.

We thus conclude that the expected node degrees are *invariant* with respect to the distribution characterizing the propagation effects.

### E. Effect of the Secrecy Rate Threshold and Noise Powers

In this section, we study the effect of non-zero secrecy rate threshold, i.e., $\varrho > 0$, and unequal noise powers, i.e., $\sigma_\ell^2 \neq \sigma_e^2$, on the $i\mathcal{S}$-graph. Considering $Z_{x_i,x_j} = 1$ and arbitrary noise powers $\sigma_\ell^2, \sigma_e^2$, we can combine (2) with the general propagation model of (1) and obtain the edge set

$$\mathcal{E} = \left\{\overrightarrow{x_i x_j} : g(|x_i - x_j|) > \frac{\sigma_\ell^2}{\sigma_e^2} 2^\varrho g(|x_i - e^*|) + \frac{\sigma_\ell^2}{P_\ell}(2^\varrho - 1)\right\}, \quad (16)$$

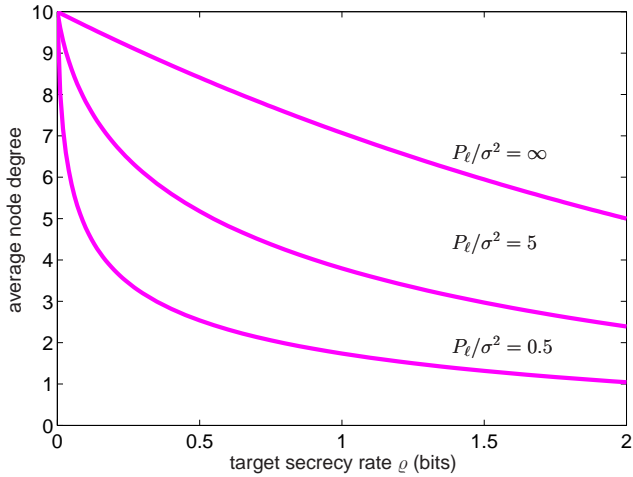
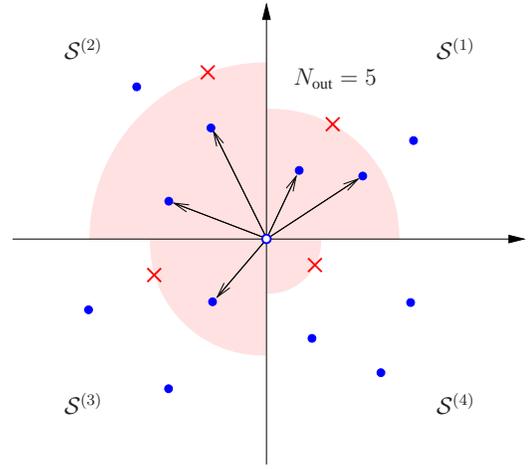

Figure 6. Average node degree versus the secrecy rate threshold $\varrho$, for various values of $P_\ell/\sigma^2$ ($\sigma_\ell^2 = \sigma_e^2 = \sigma^2$, $g(r) = \frac{1}{r^{2b}}$, $b = 2$, $\lambda_\ell = 1\,\mathrm{m}^{-2}$, $\lambda_e = 0.1\,\mathrm{m}^{-2}$).

Figure 7. Secure communication with sectorized transmission. In this example with $L = 4$ sectors, the node at the origin can transmit messages with information-theoretic security to $N_\mathrm{out} = 5$ nodes.

where

$$e^* = \underset{e_k \in \Pi_e}{\operatorname{argmin}} |x_i - e_k|.$$

Note that by setting $\varrho = 0$ and $\sigma_\ell^2 = \sigma_e^2$ in (16) we obtain the edge set in (4) as a special case. The exact dependence of the average node degree on the parameters $\varrho, \sigma_\ell^2, \sigma_e^2$ depends on the function $g(r)$. To gain further insights, we consider the specific channel gain function

$$g(r) = \frac{1}{r^{2b}}, \quad r > 0. \tag{17}$$

This function has been widely used in the literature to model path loss behavior as a function of distance, and satisfies the conditions in Section II-A. The following theorem characterizes the average degrees of the resulting $i\mathcal{S}$-graph.

*Theorem 3.4 ([32]):* For the Poisson $i\mathcal{S}$-graph with secrecy rate threshold $\varrho$, noise powers $\sigma_\ell^2, \sigma_e^2$, and channel gain function $g(r) = \frac{1}{r^{2b}}$, the average node degrees are

$$\mathbb{E}\{N_\mathrm{in}\} = \mathbb{E}\{N_\mathrm{out}\}$$
$$= \pi^2 \lambda_\ell \lambda_e \int_0^\infty \frac{x e^{-\pi \lambda_e x}}{\left(\frac{\sigma_\ell^2}{\sigma_e^2} 2^\varrho + \frac{\sigma_\ell^2}{P_\ell}(2^\varrho - 1)x^b\right)^{1/b}} dx \tag{18}$$

Figure 6 illustrates the effect of the secrecy rate threshold $\varrho$ on the average node degrees. We observe that the average node degree attains its maximum value of $\frac{\lambda_\ell}{\lambda_e} = 10$ at $\varrho = 0$, and is monotonically decreasing with $\varrho$.

## IV. TECHNIQUES FOR COMMUNICATION WITH ENHANCED SECRECY

Based on the results derived in Section III, we observe that even a small density of eavesdroppers is enough to significantly disrupt connectivity of the $i\mathcal{S}$-graph. For example, if the density of eavesdroppers is half the density of legitimate nodes, then from (12) the average node degree is only $\frac{\lambda_\ell}{\lambda_e} = 2$. In this section, we propose two techniques—*sectorized transmission* and *eavesdropper neutralization*—which achieve an average degree higher than $\frac{\lambda_\ell}{\lambda_e}$.

### A. Sectorized Transmission

We have so far assumed that the legitimate nodes employ omnidirectional antennas, distributing power equally among all directions. We now consider that each legitimate node is able to transmit independently in $L$ sectors of the plane, with $L \geq 1$, as depicted in Figure 7. This can be accomplished, for example, through the use of $L$ directional antennas. With each node $x_i \in \Pi_\ell$, we associate $L$ transmission sectors $\{\mathcal{S}_i^{(l)}\}_{l=1}^L$, defined as

$$\mathcal{S}_i^{(l)} \triangleq \left\{ z \in \mathbb{R}^2 : \phi_i + (l-1)\frac{2\pi}{L} < \angle \overrightarrow{x_i z} < \phi_i + l\frac{2\pi}{L} \right\}$$

for $l = 1 \ldots L$, where $\{\phi_i\}_{i=1}^\infty$ are random offset angles with an arbitrary joint distribution. The resulting $i\mathcal{S}$-graph $G_L = \{\Pi_\ell, \mathcal{E}_L\}$ has an edge set given by

$$\mathcal{E}_L = \{\overrightarrow{x_i x_j} : |x_i - x_j| < |x_i - e^*|\},$$

where

$$e^* = \underset{e_k \in \Pi_e \cap \mathcal{S}^*}{\operatorname{argmin}} |x_i - e_k|, \qquad \mathcal{S}^* = \left\{ \mathcal{S}_i^{(l)} : x_j \in \mathcal{S}_i^{(l)} \right\}.$$

Here, $\mathcal{S}^*$ is the transmission sector of $x_i$ that contains the destination node $x_j$, and $e^*$ is the eavesdropper inside $\mathcal{S}^*$ that is closest to the transmitter $x_i$. The following theorem characterizes the average node degree as a function of $L$.

*Theorem 4.1 (Sectorized Transmission [44]):* For the Poisson $i\mathcal{S}$-graph $G_L$ with $L$ sectors, the average node degrees are

$$\mathbb{E}\{N_\mathrm{in}\} = \mathbb{E}\{N_\mathrm{out}\} = L\frac{\lambda_\ell}{\lambda_e}. \tag{19}$$

We conclude that the average node degree increases *linearly* with the number of sectors $L$, and hence sectorized transmission is an effective technique for enhancing the secrecy of communications. Figure 7 provides an intuitive understanding of why sectorization works. Specifically, if there was no sectorization, node 0 would be out-isolated, due to the close proximity of the eavesdropper in sector $\mathcal{S}^{(4)}$. However, if we allow independent transmissions in four non-overlapping sectors, that same eavesdropper can only hear the transmissions inside sector $\mathcal{S}^{(4)}$. Thus, even though node 0 is out-isolated with respect to sector $\mathcal{S}^{(4)}$, it may still communicate securely with some legitimate nodes inside sectors $\mathcal{S}^{(1)}$, $\mathcal{S}^{(2)}$, and $\mathcal{S}^{(3)}$.

### B. Eavesdropper Neutralization

In some scenarios, each legitimate node may be able to physically inspect its surroundings and deactivate the eavesdroppers falling inside some neutralization region. With each node $x_i \in \Pi_\ell$, we associate a *neutralization region* $\Theta_i$ inside which all eavesdroppers have been deactivated. The *total neutralization region* $\Theta$ can then be seen as a Boolean model with points $\{x_i\}$ and associated sets $\{\Theta_i\}$, i.e.,

$$\Theta = \bigcup_{i=1}^{\infty}(x_i + \Theta_i).$$

Since the eavesdroppers inside $\Theta$ have been deactivated, the *effective eavesdropper process* after neutralization is $\Pi_e \cap \overline{\Theta}$, where $\overline{\Theta} \triangleq \mathbb{R}^2 \setminus \Theta$ denotes the complement of $\Theta$. The resulting $i\mathcal{S}$-graph $G_\Theta = \{\Pi_\ell, \mathcal{E}_\Theta\}$ has an edge set given by

$$\mathcal{E}_\Theta = \left\{ \overrightarrow{x_i x_j} : |x_i - x_j| < |x_i - e^*|, \quad e^* = \underset{e_k \in \Pi_e \cap \overline{\Theta}}{\operatorname{argmin}} |x_i - e_k| \right\}.$$

In the following, we consider the case of a circular neutralization set, i.e, $\Theta_i = \mathcal{B}_0(\rho)$ where $\rho$ is a deterministic *neutralization radius*, as depicted in Fig. 8.

*Theorem 4.2 (Eavesdropper Neutralization [44]):* For the enhanced Poisson $i\mathcal{S}$-graph $G_\rho$ with neutralization radius $\rho$, the average node degrees of a typical node are lower-bounded by

$$\mathbb{E}\{N_{\text{in}}\} = \mathbb{E}\{N_{\text{out}}\} \geq \frac{\lambda_\ell}{\lambda_e} \left( \pi \lambda_e \rho^2 + e^{\pi \lambda_\ell \rho^2} \right). \quad (20)$$

We conclude that the average node degree increases at a rate that is at least *exponential* with the neutralization radius $\rho$, making eavesdropper neutralization an effective technique for enhancing the secrecy of communications.

Figure 9 plots the average node degree versus the neutralization radius $\rho$, for various values of $\lambda_e$. We observe that the analytical lower-bound for $\mathbb{E}\{N_{\text{out}}\}$ given in (20) is very close to the actual value of $\mathbb{E}\{N_{\text{out}}\}$ obtained through Monte Carlo simulation, becoming tight as $\rho \to 0$ or $\lambda_e \to \infty$.

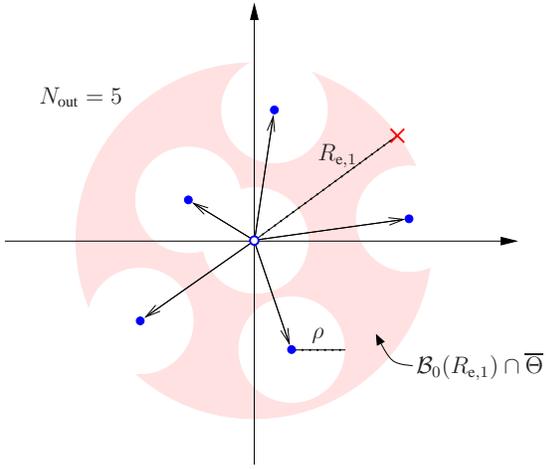

Figure 8. Secure communication with eavesdropper neutralization. In this example, the node at the origin can transmit messages with information-theoretic security to $N_{\text{out}} = 5$ nodes.

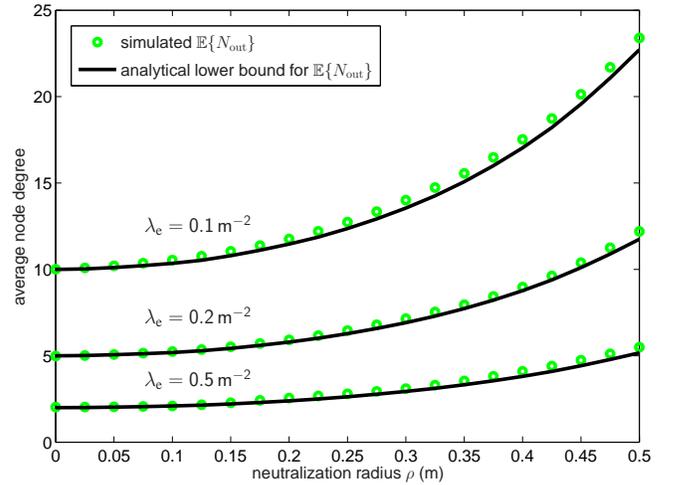

Figure 9. Average node degree versus the neutralization radius $\rho$, for various values of $\lambda_e$ ($\lambda_\ell = 1\,\text{m}^{-2}$).

## V. MAXIMUM SECRECY RATE IN THE POISSON $i\mathcal{S}$-GRAPH

In this section, we analyze the MSR between a node and each of its neighbours. Considering the coordinate system depicted in Fig. 3 and the channel gain $g(r) = \frac{1}{r^{2b}}$, the MSR $\mathcal{R}_{\mathsf{s},i}$ between the node at the origin and its $i$-th closest neighbour, $i \geq 1$, can be written for a given realization of the node positions $\Pi_\ell$ and $\Pi_e$ as

$$\mathcal{R}_{\mathsf{s},i} = \left[ \log_2 \left( 1 + \frac{P_\ell}{R_{\ell,i}^{2b} \sigma^2} \right) - \log_2 \left( 1 + \frac{P_\ell}{R_{\text{e},1}^{2b} \sigma^2} \right) \right]^+ \quad (21)$$

in bits per complex dimension. For each instantiation of the random Poisson processes $\Pi_\ell$ and $\Pi_e$, a realization of the RV $\mathcal{R}_{\mathsf{s},i}$ is obtained. The following theorem provides the distribution of this RV.

*Theorem 5.1 ([31]):* The MSR $\mathcal{R}_{\mathsf{s},i}$ between a typical node and its $i$-th closest neighbour, $i \geq 1$, is a RV whose cumulative distribution function (CDF) $F_{\mathcal{R}_{\mathsf{s},i}}(\varrho)$ is given by

$$F_{\mathcal{R}_{\mathsf{s},i}}(\varrho) = 1 - \frac{\ln 2 (\pi \lambda_\ell)^i}{(i-1)! b} \left( \frac{P_\ell}{\sigma^2} \right)^{\frac{i}{b}} \int_\varrho^{+\infty} \frac{2^z}{(2^z - 1)^{1+\frac{i}{b}}}$$
$$\times \exp\left( -\pi \lambda_\ell \left( \frac{\frac{P_\ell}{\sigma^2}}{2^z - 1} \right)^{\frac{1}{b}} - \pi \lambda_e \left( \frac{\frac{P_\ell}{\sigma^2}}{2^{z-\varrho} - 1} \right)^{\frac{1}{b}} \right) dz, \quad (22)$$

for $\varrho \geq 0$.

From this result, we can trivially obtain the probability of existence of a non-zero MSR, and the probability of secrecy outage.

*Corollary 5.1:* Considering the link between a typical node and its $i$-th closest neighbour, $i \geq 1$, the probability of *existence* of a non-zero MSR, $p_{\text{exist},i} = \mathbb{P}\{\mathcal{R}_{\mathsf{s},i} > 0\}$, is

$$p_{\text{exist},i} = \left( \frac{\lambda_\ell}{\lambda_\ell + \lambda_e} \right)^i \quad (23)$$

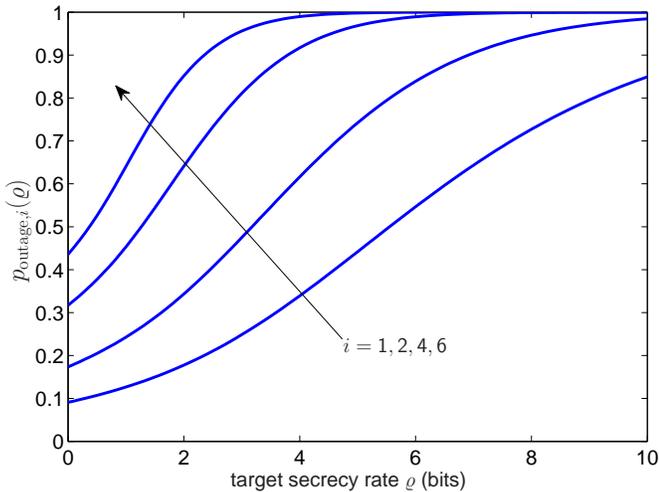
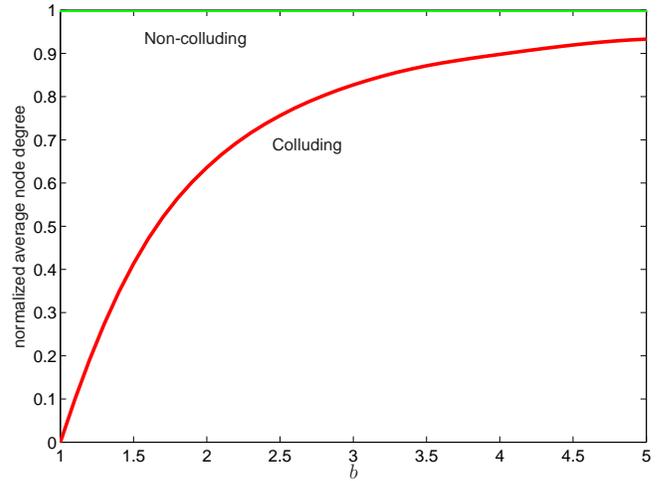

Figure 10. Probability $p_{\text{outage},i}$ of secrecy outage between a node and its $i$-th closest neighbour, for various values of the neighbour index $i$ ($\lambda_\ell = 1\,\text{m}^{-2}$, $\lambda_e = 0.1\,\text{m}^{-2}$, $b = 2$, $P_\ell/\sigma^2 = 10$).

Figure 11. Normalized average node degree of the $i\mathcal{S}$-graph, $\frac{\mathbb{E}\{N_{\text{out}}\}}{\lambda_\ell/\lambda_e}$, versus the amplitude loss exponent $b$, for the cases of colluding and non-colluding eavesdroppers.

and the probability of an *outage* in MSR is $p_{\text{outage},i}(\varrho) = \mathbb{P}\{\mathcal{R}_{\mathsf{s},i} < \varrho\} = F_{\mathcal{R}_{\mathsf{s},i}}(\varrho)$, as given in (22).

Figure 10 shows the probability $p_{\text{outage},i}$ of secrecy outage of a typical node transmitting to its $i$-th neighbour, as a function of the desired secrecy rate $\varrho$. As expected, a secrecy outage become more likely as we increase the target secrecy rate $\varrho$ set by the transmitter.

## VI. THE CASE OF COLLUDING EAVESDROPPERS

In this section, we consider that the eavesdroppers have ability to *collude*, i.e., they can exchange and combine the information received by all the eavesdroppers to decode the secret message. The following theorem characterizes the resulting average node degree in such graph.

*Theorem 6.1 ([32]):* For the Poisson $i\mathcal{S}$-graph with colluding eavesdroppers, secrecy rate threshold $\varrho = 0$, equal noise powers $\sigma_\ell^2 = \sigma_e^2$, and channel gain function $g(r) = \frac{1}{r^{2b}}$, $b > 1$, the average degrees of a typical node are

$$\mathbb{E}\{N_{\text{in}}\} = \mathbb{E}\{N_{\text{out}}\} = \frac{\lambda_\ell}{\lambda_e}\,\text{sinc}\left(\frac{1}{b}\right), \qquad (24)$$

where $\text{sinc}(x) \triangleq \frac{\sin(\pi x)}{\pi x}$.

It is insightful to rewrite (24) as $\mathbb{E}\{N_{\text{out}}|\text{colluding}\} = \mathbb{E}\{N_{\text{out}}|\text{non-colluding}\} \cdot \eta(b)$, where $\eta(b) = \text{sinc}(\frac{1}{b})$, and $\eta(b) < 1$ for $b > 1$. The function $\eta(b)$ can be interpreted as the *degradation factor in average connectivity due to eavesdropper collusion*. In the extreme where $b = 1$, we have complete loss of secure connectivity with $\eta(1) = 0$. This is because the series $P_{\text{rx,e}} = \sum_{i=1}^{\infty} \frac{P_\ell}{R_{e,i}^{2b}}$ diverges (i.e., the total received eavesdropper power is infinite), so the resulting average node degree is zero. In the other extreme where $b \to \infty$, we achieve the highest secure connectivity with $\eta(\infty) = 1$. This is because the first term $\frac{P_\ell}{R_{e,1}^{2b}}$ in the $P_{\text{rx,e}}$ series (corresponding to the non-colluding term) is dominant, so the average node degree in the colluding case approaches the non-colluding one.

Figure 11 quantifies the (normalized) average node degree of the $i\mathcal{S}$-graph, $\frac{\mathbb{E}\{N_{\text{out}}\}}{\lambda_\ell/\lambda_e}$, versus the amplitude loss exponent $b$. As predicted analytically, it is apparent that cluttered environments with larger amplitude loss exponents $b$ are more favorable for secure communication, in the sense that in such environments collusion only provides a marginal performance improvement for the eavesdroppers.

## VII. PERCOLATION IN THE POISSON $i\mathcal{S}$-GRAPH

Percolation theory studies the existence of phase transitions in random graphs, whereby an infinite cluster of connected nodes suddenly arises as some system parameter is varied. Percolation theory has been used to study connectivity of multi-hop wireless networks, where the formation of an unbounded cluster is desirable for communication over arbitrarily long distances [45]–[47]. In this section, we prove the existence of a phase transition in the Poisson $i\mathcal{S}$-graph, showing that long-range communication in a wireless network is still possible when a secrecy constraint is present.

### A. Definitions

*Graphs:* We use $G = \{\Pi_\ell, \mathcal{E}\}$ to denote the (directed) $i\mathcal{S}$-graph with vertex set $\Pi_\ell$ and edge set given in (2). In addition, we define two undirected graphs: the *weak $i\mathcal{S}$-graph* $G^{\text{weak}} = \{\Pi_\ell, \mathcal{E}^{\text{weak}}\}$, where

$$\mathcal{E}^{\text{weak}} = \{\overline{x_i x_j} : \mathcal{R}_{\mathsf{s}}(x_i, x_j) > \varrho \vee \mathcal{R}_{\mathsf{s}}(x_j, x_i) > \varrho\},$$

and the *strong $i\mathcal{S}$-graph* $G^{\text{strong}} = \{\Pi_\ell, \mathcal{E}^{\text{strong}}\}$, where

$$\mathcal{E}^{\text{strong}} = \{\overline{x_i x_j} : \mathcal{R}_{\mathsf{s}}(x_i, x_j) > \varrho \wedge \mathcal{R}_{\mathsf{s}}(x_j, x_i) > \varrho\}.$$

*Graph Components:* We use the notation $x \xrightarrow{G} y$ to represent a path from node $x$ to node $y$ in a directed graph $G$, and $x \xleftrightarrow{G^*} y$ to represent a path between node $x$ and node $y$ in an undirected

graph $G^*$. We define four components:

$$\mathcal{K}^{\text{out}}(x) \triangleq \{y \in \Pi_\ell : \exists x \xrightarrow{G} y\}, \quad (25)$$

$$\mathcal{K}^{\text{in}}(x) \triangleq \{y \in \Pi_\ell : \exists y \xrightarrow{G} x\}, \quad (26)$$

$$\mathcal{K}^{\text{weak}}(x) \triangleq \{y \in \Pi_\ell : \exists x \overset{G^{\text{weak}}}{—} y\}, \quad (27)$$

$$\mathcal{K}^{\text{strong}}(x) \triangleq \{y \in \Pi_\ell : \exists x \overset{G^{\text{strong}}}{—} y\}. \quad (28)$$

*Percolation Probabilities:* To study percolation in the $i\mathcal{S}$-graph, it is useful to define percolation probabilities associated with the four graph components. Specifically, let $p_\infty^{\text{out}}$, $p_\infty^{\text{in}}$, $p_\infty^{\text{weak}}$, and $p_\infty^{\text{strong}}$ respectively be the probabilities that the in, out, weak, and strong components containing node $x = 0$ have an infinite number of nodes, i.e.,

$$p_\infty^\diamond(\lambda_\ell, \lambda_e, \varrho) \triangleq \mathbb{P}\{|\mathcal{K}^\diamond(0)| = \infty\}$$

for $\diamond \in \{\text{out}, \text{in}, \text{weak}, \text{strong}\}$.[5]

### B. Main Result

Typically, a continuum percolation model consists of an underlying point process defined on the infinite plane, and a rule that describes how connections are established between the nodes [48]. A main property of all percolation models is that they exhibit a *phase transition* as some continuous parameter is varied. If this parameter is the density $\lambda$ of nodes, then the phase transition occurs at some *critical density* $\lambda_c$. When $\lambda < \lambda_c$, denoted as the *subcritical phase*, all the clusters are a.s. bounded.[6] When $\lambda > \lambda_c$, denoted as the *supercritical phase*, the graph exhibits a.s. an unbounded cluster of nodes, or in other words, the graph *percolates*.

We now determine if percolation in the $i\mathcal{S}$-graph is possible, and if so, for which combinations of system parameters $(\lambda_\ell, \lambda_e, \varrho)$ does it occur. The mathematical characterization of the $i\mathcal{S}$-graph presents two challenges: i) the $i\mathcal{S}$-graph is a directed graph, which leads to the study of *directed percolation*; and ii) the $i\mathcal{S}$-graph exhibits dependencies between the state of different edges, which leads to the study of *dependent percolation*. The result is given by the following main theorem.

*Theorem 7.1 (Phase Transition in the $i\mathcal{S}$-Graph [35]):*
For any $\lambda_e > 0$ and $\varrho$ satisfying

$$0 \leq \varrho < \varrho_{\max} \triangleq \log_2\left(1 + \frac{P \cdot g(0)}{\sigma^2}\right), \quad (29)$$

there exist critical densities $\lambda_c^{\text{out}}$, $\lambda_c^{\text{in}}$, $\lambda_c^{\text{weak}}$, $\lambda_c^{\text{strong}}$ satisfying

$$0 < \lambda_c^{\text{weak}} \leq \lambda_c^{\text{out}} \leq \lambda_c^{\text{strong}} < \infty \quad (30)$$

$$0 < \lambda_c^{\text{weak}} \leq \lambda_c^{\text{in}} \leq \lambda_c^{\text{strong}} < \infty \quad (31)$$

such that

$$p_\infty^\diamond = 0, \quad \text{for } \lambda_\ell < \lambda_c^\diamond, \quad (32)$$

$$p_\infty^\diamond > 0, \quad \text{for } \lambda_\ell > \lambda_c^\diamond, \quad (33)$$

for any $\diamond \in \{\text{out}, \text{in}, \text{weak}, \text{strong}\}$. Conversely, if $\varrho > \varrho_{\max}$, then $p_\infty^\diamond = 0$ for any $\lambda_\ell, \lambda_e$.

---

[5]Except where otherwise indicated, we use the symbol $\diamond$ to represent the out, in, weak, or strong component.

[6]We say that an event occurs "almost surely" (a.s.) if its probability is equal to one.

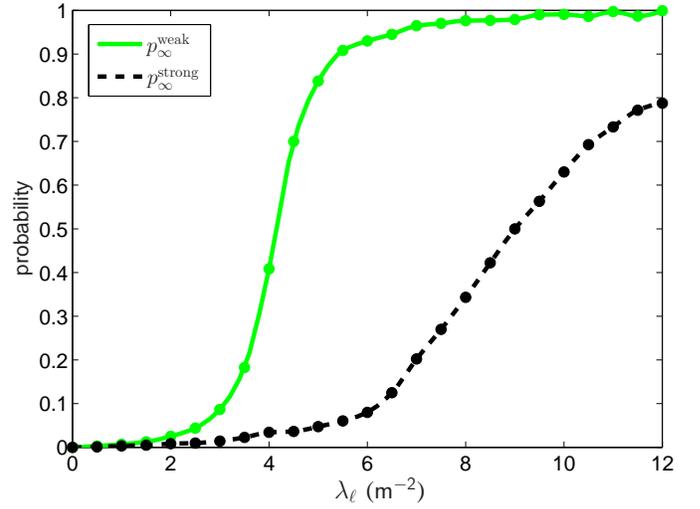

Figure 12. Simulated percolation probabilities for the weak and strong components of the $i\mathcal{S}$-graph, versus the density $\lambda_\ell$ of legitimate nodes ($\lambda_e = 1 \, \text{m}^{-2}$, $\varrho = 0$).

Theorem 7.1 shows that each of the four components of the $i\mathcal{S}$-graph (in, out, weak, and strong) experiences a phase transition at some nontrivial critical density $\lambda_c^\diamond$ of legitimate nodes. In addition, it shows that percolation can occur for any prescribed secrecy threshold $\varrho$ satisfying $\varrho < \varrho_{\max} = \log_2\left(1 + \frac{P \cdot g(0)}{\sigma^2}\right)$, as long as the density of legitimate nodes is made large enough. This implies that for unbounded path loss models such as $g(r) = 1/r^\gamma$, percolation can occur for *any* arbitrarily large secrecy requirement $\varrho$, while for bounded models such as $g(r) = 1/(1 + r^\gamma)$, the desired $\varrho$ may be too high to allow percolation. Our results also show that as long as $\varrho < \varrho_{\max}$, percolation can be achieved even in cases where the eavesdroppers are arbitrarily dense, by making the density of legitimate nodes large enough.

Figure 12 shows the percolation probabilities for the weak and strong components of the $i\mathcal{S}$-graph, versus the density $\lambda_\ell$ of legitimate nodes. As predicted by Theorem 7.1, the figure suggests that these components experience a phase transition as $\lambda_\ell$ is increased. In particular, $\lambda_c^{\text{weak}} \approx 3.4 \, \text{m}^{-2}$ and $\lambda_c^{\text{strong}} \approx 6.2 \, \text{m}^{-2}$, for the case of $\lambda_e = 1 \, \text{m}^{-2}$ and $\varrho = 0$. Operationally, this means that if long-range bidirectional secure communication is desired in a wireless network, the density of legitimate nodes must be at least 6.2 times that of the eavesdroppers. In practice, the density of legitimate nodes must be even larger, because a secrecy requirement greater than $\varrho = 0$ is typically required. This dependence on $\varrho$ is illustrated in Figure 13. In practice, it might also be of interest to increase $\lambda_\ell$ fairly beyond the critical density, since this leads to an increased average fraction $p_\infty^\diamond$ of nodes which belong to the infinite component, thus improving secure connectivity.

## VIII. FULL CONNECTIVITY IN THE POISSON $i\mathcal{S}$-GRAPH

In the previous section, we studied percolation in the $i\mathcal{S}$-graph defined over the infinite plane. We showed that for some combinations of the parameters $(\lambda_\ell, \lambda_e, \varrho)$, the regime is supercritical and an infinite component arises. However, the existence of an infinite component does not ensure connectivity between any two nodes, e.g., one node inside the infinite component cannot communicate with a node outside. In this sense, percolation ensures only *partial connectivity* of the network. In some scenarios, it is of interest to guarantee *full connectivity*, i.e., that all nodes can communicate with each other, possibly through multiple hops. Note, however, that for networks defined over an infinite region, the probability of full connectivity is exactly zero. Thus, to study of full connectivity, we need to restrict our attention to a finite region $\mathcal{R}$. Throughout this section, we consider the simplest case of $\varrho = 0$, i.e., the *existence* of secure links with a positive (but possibly small) MSR.

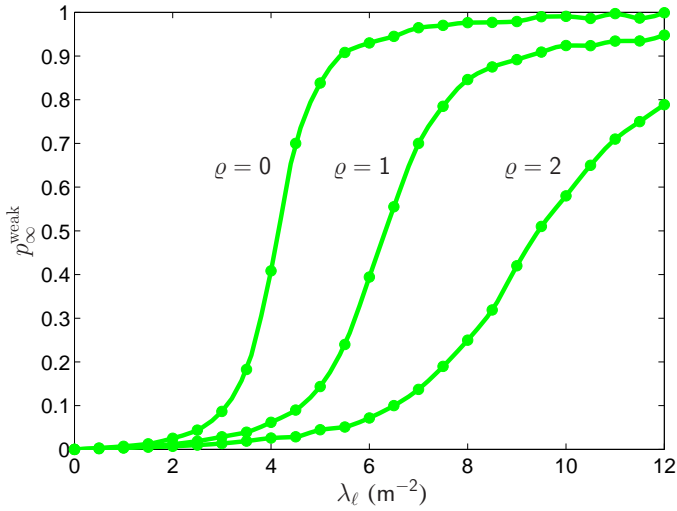

Figure 13. Effect of the secrecy rate threshold $\varrho$ on the percolation probability $p_\infty^{\text{weak}}$ ($\lambda_e = 1\,\text{m}^{-2}$, $g(r) = 1/r^4$, $P_\ell/\sigma^2 = 10$).

### A. Definitions

Since the $i\mathcal{S}$-graph is a directed graph, we start by distinguishing between full out- and in-connectivity with the following definitions.

*Definition 8.1 (Full Out-Connectivity):* A legitimate node $x_i \in \Pi_\ell \cap \mathcal{R}$ is *fully out-connected* with respect to a region $\mathcal{R}$ if in the $i\mathcal{S}$-graph $G = \{\Pi_\ell, \mathcal{E}\}$ there exists a directed path between $x_i$ and *every* node $x_j \in \Pi_\ell \cap \mathcal{R}$, for $x_i \neq x_j$.

*Definition 8.2 (Full In-Connectivity):* A legitimate node $x_i \in \Pi_\ell \cap \mathcal{R}$ is *fully in-connected* with respect to a region $\mathcal{R}$ if in the $i\mathcal{S}$-graph $G = \{\Pi_\ell, \mathcal{E}\}$ there exists a directed path between *every* node $x_j \in \Pi_\ell \cap \mathcal{R}$ and $x_i$, for $x_i \neq x_j$.

The $i\mathcal{S}$-graph is a random graph, and therefore we can consider the probabilities of a node $x_i$ being fully in- or out-connected. For analysis purposes, we consider that a probe legitimate node (node 0) placed at the origin of the coordinate system, i.e., $x_{\text{probe}} = 0 \subset \mathcal{R}$. We then define $p_{\text{out-con}}$ and $p_{\text{in-con}}$ as the probability that node $0$ is, respectively, fully out- and fully in-connected. These probabilities are a deterministic function of the densities $\lambda_\ell$ and $\lambda_e$, and the area $A$ of region $\mathcal{R}$. Our goal is to characterize $p_{\text{out-con}}$ and $p_{\text{in-con}}$.

### B. Main Result

In what follows, we focus on the asymptotic behavior of secure connectivity in the limit of a large density of legitimate nodes.[7] Specifically, for a fixed region of area $A$ and a fixed density $\lambda_e$ of eavesdroppers, we would like to determine if by increasing $\lambda_\ell \to \infty$, we can asymptotically achieve full in- and out-connectivity with probability $1$.[8] The following theorem characterizes the asymptotic out-connectivity in the $i\mathcal{S}$-graph.

*Theorem 8.1 (Asymptotic Out-Connectivity [49]):* For the Poisson $i\mathcal{S}$-graph with $\lambda_e > 0$ and $A > 0$, we have

$$\lim_{\lambda_\ell \to \infty} p_{\text{out-con}} = 1,$$

i.e., the legitimate node at the origin is asymptotically out-connected.

The following theorem characterizes the asymptotic in-connectivity in the $i\mathcal{S}$-graph.

*Theorem 8.2 (Asymptotic In-Connectivity [49]):* For the Poisson $i\mathcal{S}$-graph with $\lambda_e > 0$ and $A > 0$, we have

$$\lim_{\lambda_\ell \to \infty} p_{\text{in-con}} \leq 1 - \frac{6\pi}{8\pi + 3\sqrt{3}}(1 - e^{-\lambda_e A}), \quad (34)$$

i.e., the legitimate node at the origin is *not* asymptotically in-connected.

The theorems show that full out-connectivity can be improved as much as desired by deploying more legitimate nodes. Full in-connectivity, however, remains bounded away from 1, no matter how large $\lambda_\ell$ is made (an intuitive explanation for this fact is provided in [49]). Operationally, this means that a node can a.a.s. *transmit* secret messages to all the nodes in a finite region $\mathcal{R}$, but cannot a.a.s. *receive* secret messages from all the nodes in $\mathcal{R}$.

## IX. CONCLUSION

Using the notion of strong secrecy, we provided an information-theoretic definition of the $i\mathcal{S}$-graph as a model for intrinsically secure communication in large-scale networks. The $i\mathcal{S}$-graph captures the connections that can be established with MSR exceeding a threshold $\varrho$, in a large network. This paper provided an overview of the main properties of this new class of random graphs.

Perhaps the most interesting insight to be gained from our results, is the exact quantification of the impact of the eavesdropper density $\lambda_e$ on secure connectivity—a modest density of scattered eavesdroppers can potentially cause a drastic reduction in the MSR provided at the physical layer of wireless communication networks. Our work has not yet addressed all of the far reaching implications of the broadcast property of the wireless medium. In the most general scenario, legitimate nodes could for example transmit their signals in a cooperative fashion, whereas malicious nodes could use jamming to disrupt all communications. We hope that further efforts in combining stochastic geometry with information-theoretic principles will lead to a more comprehensive treatment of wireless security.

---

[7]For a non-asymptotic analysis of secure full connectivity, see [49].
[8]We say that an event occurs "asymptotically almost surely" (a.a.s.) if its probability approaches one as $\lambda_\ell \to \infty$.

ACKNOWLEDGEMENTS

The authors would like to thank L. A. Shepp, J. N. Tsitsiklis, V. K. Goyal, Y. Shen, and W. Suwansantisuk for their helpful suggestions.

## REFERENCES

[1] C. E. Shannon, "Communication theory of secrecy systems," *Bell System Technical Journal*, vol. 29, pp. 656–715, 1949.

[2] A. D. Wyner, "The Wire-Tap Channel," *Bell System Technical Journal*, vol. 54, no. 8, pp. 1355–1367, October 1975.

[3] I. Csiszár and J. Korner, "Broadcast channels with confidential messages," *IEEE Trans. Inf. Theory*, vol. 24, no. 3, pp. 339–348, 1978.

[4] S. Leung-Yan-Cheong and M. Hellman, "The Gaussian wire-tap channel," *IEEE Trans. Inf. Theory*, vol. 24, no. 4, pp. 451–456, July 1978.

[5] A. Hero, "Secure space-time communication," *IEEE Transactions on Information*, vol. 49, no. 12, pp. 3235–3249, Dec. 2003.

[6] R. Negi and S. Goel, "Secret communication using artificial noise," in *Proc. IEEE Vehicular Technology Conference*, vol. 3, Dallas, TX, Sept. 2005, pp. 1906–1910.

[7] E. Ekrem and S. Ulukus, "Secrecy capacity region of the gaussian multi-receiver wiretap channel," in *Proc. IEEE Int. Symp. on Inf. Theory*, Seoul, Korea, June 2009, pp. 2612–2616.

[8] T. Liu and S. Shamai, "A note on the secrecy capacity of the multiple-antenna wiretap channel," *IEEE Trans. Inf. Theory*, vol. 55, no. 6, pp. 2547–2553, June 2009.

[9] H. Weingarten, T. Liu, S. Shamai, Y. Steinberg, and P. Viswanath, "The capacity region of the degraded multiple-input multiple-output compound broadcast channel," *IEEE Trans. Inf. Theory*, vol. 55, no. 11, pp. 5011–5023, Nov. 2009.

[10] L. Zhang, R. Zhang, Y. Liang, Y. Xin, and S. Cui, "On the relationship between the multi-antenna secrecy communications and cognitive radio communications," in *Proc. Allerton Conf. on Communications, Control, and Computing*, Monticello, IL, Sept. 2009.

[11] S. Goel and R. Negi, "Secret communication in presence of colluding eavesdroppers," in *Proc. Military Commun. Conf.*, Oct. 2005, pp. 1501–1506.

[12] P. C. Pinto, J. O. Barros, and M. Z. Win, "Wireless physical-layer security: The case of colluding eavesdroppers," in *Proc. IEEE Int. Symp. on Inf. Theory*, Seoul, South Korea, July 2009, pp. 2442–2446.

[13] E. Ekrem and S. Ulukus, "Secrecy in cooperative relay broadcast channels," in *Proc. IEEE Int. Symp. on Inf. Theory*, Toronto, ON, July 2008, pp. 2217–2221.

[14] P. Parada and R. Blahut, "Secrecy capacity of SIMO and slow fading channels," in *Proc. IEEE Int. Symp. on Inf. Theory*, Adelaide, Australia, Sept. 2005, pp. 2152–2155.

[15] P. Gopala, L. Lai, and H. El Gamal, "On the Secrecy Capacity of Fading Channels," arxiv preprint cs.IT/0610103, 2006.

[16] Z. Li, R. Yates, and W. Trappe, "Secrecy capacity of independent parallel channels," *Proc. Annu. Allerton Conf. Communication, Control and Computing*, pp. 841–848, Sept. 2006.

[17] M. Bloch, J. Barros, M. R. D. Rodrigues, and S. W. McLaughlin, "Wireless information-theoretic security," *IEEE Trans. Inf. Theory*, vol. 54, no. 6, pp. 2515–2534, 2008.

[18] Y. Liang, H. V. Poor, and S. Shamai, "Secure communication over fading channels," *IEEE Trans. Inf. Theory*, vol. 54, pp. 2470–2492, June 2008.

[19] U. Maurer and S. Wolf, "Information-theoretic key agreement: From weak to strong secrecy for free," *Eurocrypt 2000, Lecture Notes in Computer Science*, vol. 1807, pp. 351+, 2000.

[20] J. Barros and M. Bloch, "Strong secrecy for wireless channels," in *Proc. International Conf. on Inf. Theor. Security*, Calgary, Canada, Aug. 2008.

[21] U. Maurer, "Secret key agreement by public discussion from common information," *IEEE Trans. Inf. Theory*, vol. 39, no. 3, pp. 733–742, May 1993.

[22] R. Ahlswede and I. Csiszar, "Common randomness in information theory and cryptography - Part I: Secret sharing," *IEEE Trans. Inf. Theory*, vol. 39, no. 4, pp. 1121–1132, July 1993.

[23] C. H. Bennett, G. Brassard, C. Crépeau, and U. Maurer, "Generalized privacy amplification," *IEEE Trans. Inf. Theory*, vol. 41, no. 6, pp. 1915–1923, 1995.

[24] U. M. Maurer and S. Wolf, "Unconditionally secure key agreement and intrinsic conditional information," *IEEE Trans. Inf. Theory*, vol. 45, no. 2, pp. 499–514, March 1999.

[25] M. Bloch, A. Thangaraj, S. W. McLaughlin, and J.-M. Merolla, "LDPC-based secret key agreement over the gaussian wiretap channel," in *Proc. IEEE Int. Symp. on Inf. Theory*, Seattle, USA, 2006.

[26] A. Thangaraj, S. Dihidar, A. R. Calderbank, S. W. McLaughlin, and J.-M. Merolla, "Applications of LDPC codes to the wiretap channel," *IEEE Trans. Inf. Theory*, vol. 53, no. 8, pp. 2933–2945, Aug. 2007.

[27] R. Liu, Y. Liang, H. V. Poor, and P. Spasojevic, "Secure nested codes for type II wiretap channels," in *Proc. IEEE Inf. Theory Workshop*, Tahoe City, CA, Sept. 2007, pp. 337–342.

[28] J. Muramatsu, "Secret key agreement from correlated source outputs using low density parity check matrices," *IEICE Trans. on Fund. of Elec. Comm. Comp.*, vol. E89-A, no. 7, pp. 2036–2046, July 2006.

[29] M. Bloch and J. O. Barros, *Physical-Layer Security*. Cambridge University Press, 2011.

[30] M. Haenggi, "The secrecy graph and some of its properties," in *Proc. IEEE Int. Symp. on Inf. Theory*, Toronto, Canada, July 2008.

[31] P. C. Pinto, J. O. Barros, and M. Z. Win, "Physical-layer security in stochastic wireless networks," in *Proc. IEEE Int. Conf. on Commun. Systems*, Guangzhou, China, Nov. 2008, pp. 974–979.

[32] ——, "Secure communication in stochastic wireless networks," pp. 1–59, 2009, submitted for publication, preprint available on arXiv:1001.3697.

[33] Y. Liang, H. V. Poor, and L. Ying, "Secrecy throughput of MANETs with malicious nodes," in *Proc. IEEE Int. Symp. on Inf. Theory*, June 2009, pp. 1189–1193.

[34] O. O. Koyluoglu, C. E. Koksal, and H. E. Gamal, "On secrecy capacity scaling in wireless networks," in *Inf. Theory and Applications Workshop*, San Diego, CA, Feb. 2010, pp. 1–4.

[35] P. C. Pinto and M. Z. Win, "Continuum percolation in the intrinsically secure communications graph," in *Proc. IEEE Int. Symp. on Inf. Theory and Its Applications*, Taichung, Taiwan, Oct. 2010, pp. 1–6.

[36] H. Inaltekin, M. Chiang, H. V. Poor, and S. B. Wicker, "The behavior of unbounded path-loss models and the effect of singularity on computed network characteristics," *IEEE J. Sel. Areas Commun.*, vol. 27, no. 7, pp. 1078–1092, Sept. 2009.

[37] M. Z. Win, P. C. Pinto, and L. A. Shepp, "A mathematical theory of network interference and its applications," *Proc. IEEE*, vol. 97, no. 2, pp. 205–230, Feb. 2009, special issue on *Ultra-Wide Bandwidth (UWB) Technology & Emerging Applications*.

[38] J. Kingman, *Poisson Processes*. Oxford University Press, 1993.

[39] D. Stoyan, W. S. Kendall, and J. Mecke, *Stochastic geometry and its applications*. John Wiley & Sons, 1995.

[40] M. Haenggi, J. G. Andrews, F. Baccelli, O. Dousse, and M. Franceschetti, "Stochastic geometry and random graphs for the analysis and design of wireless networks," *IEEE J. Sel. Areas Commun.*, vol. 27, no. 7, pp. 1029–1046, 2009.

[41] M. Abramowitz and I. A. Stegun, *Handbook of Mathematical Functions*. Dover Publications, 1970.

[42] E. N. Gilbert, "Random subdivisions of space into crystals," *Ann. Math. Statist.*, vol. 33, pp. 958–972, 1962.

[43] K. A. Brakke, "200,000,000 random Voronoi polygons," *unpublished*.

[44] P. C. Pinto, J. O. Barros, and M. Z. Win, "Techniques for enhanced physical-layer security," in *Proc. IEEE Global Telecomm. Conf.*, Miami, FL, Dec. 2010, pp. 1–5.

[45] E. N. Gilbert, "Random plane networks," *Journal of the Society for Industrial and Applied Mathematics*, vol. 9, p. 533, 1961.

[46] M. D. Penrose, "On a continuum percolation model," *Advances in Applied Probability*, vol. 23, no. 3, pp. 536–556, 1991.

[47] M. Franceschetti and R. Meester, *Random Networks for Communication*. Cambridge University Press, 2007.

[48] R. Meester and R. Roy, *Continuum Percolation*. Cambridge University Press, 1996.

[49] P. C. Pinto and M. Z. Win, "Full connectivity in the intrinsically secure communications graph," in *Proc. IEEE Int. Conf. on Commun.*, Kyoto, Japan, June 2011, pp. 1–6, to appear.